\pdfoutput=1

\documentclass{iopart}
\usepackage{iopams}
\usepackage[english]{babel}
\usepackage{cite} 
\usepackage{textcomp}
\usepackage{graphics}
\usepackage{graphicx}
\usepackage{anysize}
\usepackage{amssymb}
\usepackage{subfigure}
\usepackage{float}

\usepackage{array}

\newcommand{\text}[1]{{\rm #1}}

\usepackage{color}
\newcommand{\NEW}[1]{{#1}}

\begin{document}

\title{Flow and rheology of frictional elongated grains}

\author{
  D\'aniel B. Nagy,\textit{$^{1,2}$}
  Philippe Claudin,\textit{$^{2}$}
  Tam\'as B\"orzs\"onyi,\textit{$^{1}$} and
  Ell\'ak Somfai$^{\ast}$\textit{$^{1}$}
  }
	
\address{
$^1$Institute for Solid State Physics and Optics, Wigner Research Centre for
Physics, P.O. Box 49, H-1525 Budapest,\\
$^2$ Physique et M\'ecanique des Milieux H\'et\'erog\`enes,
  PMMH UMR 7636 CNRS, ESPCI PSL Research University, Sorbonne Universit\'e,
  Universit\'e de Paris, 10 rue Vauquelin, 75005 Paris, France
  }

\ead{\NEW{somfai.ellak@wigner.hu}}

\begin{abstract}
  The rheology of a 3-dimensional granular system consisting of frictional 
  elongated particles was investigated by means of discrete element model (DEM)
  calculations. A
  homogenous shear flow of frictional spherocyliders was simulated, and a
  number of rheological quantities were calculated. In the framework of the
  $\mu(I)$ rheology, the effective friction was found to be a non-monotonic
  function of the aspect ratio for interparticle friction coefficient $\mu_p
  \lesssim 0.4$, while it was an increasing function for larger $\mu_p$.
  We reveal the microscopic origin of this peculiar non-monotonic behavior.
  We show the non-trivial dependence of the velocity fluctuations on the
  dissipation regime, and trace back the behavior of the normal stress
  differences to particle-level quantities.
\end{abstract}

\vspace*{40mm}
\noindent
Published in: 
\medskip

\noindent
New J. Phys. \textbf{22} 073008 (2020)\\
\texttt{https://doi.org/10.1088/1367-2630/ab91fe} \qquad (open access)
\newpage

\section{Introduction}

Dense granular flows, with their rich phenomenology, are of great interest
for fundamental questions as well as for their relevance in applied
problems. They have been the subject of a large number of studies in the
last decades, with experimental, theoretical and numerical approaches,
see the general book\cite{andreotti-book-2013}.
Taking advantage of the constant increase of computational power, numerical
simulations of granular systems have now reached an impressive degree of
realism, allowing them for reliable predictions, even in rather sophisticated
configurations\cite{radjai-book-2011}. In an effort to go beyond the ideal
case of frictionless hard spheres, particles with various
shapes\cite{donev-science-2004, cruz-pre-2005, reddy-jfluidmech-2010,
schreck-softmatter-2010, azema-pre-2010, kyrylyuk-softmatter-2011,
cegeo-epl-2012, miskin-softmatter-2014, baule-softmatter-2014, azema-pre-2015,
mandal-physfluids-2016, marschall-pre-2018, nath-epje-2019} and surface or bulk
conditions\cite{rognon-epl-2006, sun-jfluidmech-2011, chialvo-pre-2012,
kamrin-computpartmech-2014, singh-newjphys-2015, favier-prf-2017,
koivisto-softmatter-2017, roy-newjphys-2017, berzi-softmatter-2015}
(e.g., friction, cohesion, stiffness), have been modelled and investigated.

Here, our interest is focused on the rheological behaviour of assemblies
of frictional elongated grains close to jamming. 
The fundamental question is, how large is the resistance
(i.e., the effective friction $\mu$) of the material against slow shearing, and
how this effective friction changes with grain elongation.
In such systems the shear flow induces particle rotation which leads to 
more intensive collisions between neighbouring particles than for spherical 
grains. The speed of the shear induced rotation depends on the particle orientation, 
faster rotation for particles parallel to the shear gradient and slower 
rotation for particles pointing in the flow direction, which results in 
orientational order. Both of these phenomena -- collisions due to rotation 
and orientational ordering -- affect the flowing and mechanical properties of the
system\cite{borzsonyi-prl-2012, wegner-softmatter-2012,
borzsonyi-softmatter-2013, tapia-jfluidmech-2017}.
This problem has previously only been addressed numerically in simplified situations:
either with frictionless grains \cite{nagy-pre-2017} or in a 2D system
\cite{trulsson-jfluidmech-2018}.
Those studies revealed an unexpected, peculiar behaviour: the effective
friction was found to be non-monotonic (increasing and decreasing)
with increasing aspect ratio $\alpha$ for the 3D frictionless case
\cite{nagy-pre-2017}, and this non-monotonic tendency was shown to
persist even for frictional grains in a 2D system, although only at
small values of the interparticle friction coefficient (up to around
$\mu_p=0.15$) \cite{trulsson-jfluidmech-2018}.
In a real world situation the interparticle friction is significantly larger 
($\mu_p$ is around 0.3). Moreover, a 3D system is substantially different from
the 2D case as the rotating
particles have extra degrees of freedom to evade and reduce the effect of
collisions, which is not possible in 2D. DEM simulations provide both
macroscopic and microscopic information about these processes.
An effective way is to use a pressure-imposed shear geometry, where
the constitutive laws for elongated grains followed
\cite{nagy-pre-2017,trulsson-jfluidmech-2018} the general framework of the
so-called $\mu(I)$ rheology\cite{gdrmidi-eurphysje-2004, cruz-pre-2005,
jop-nature-2006, hatano-pre-2007}. In this description
the effective friction $\mu = \tau/P$
as well as the volume fraction $\phi$ are functions of the inertial number
$I=\dot \gamma d/\sqrt{P/\rho}$, where $\tau$ is the shear stress, $P$ the
pressure, $\dot \gamma$ is the shear rate, $d$ is the grain size and $\rho$ is
the grain density.  These rheological functions also depend on the aspect
ratio $\alpha$ of the grains. In addition to the above mentioned
non-monotonic behaviour of the effective friction on $\alpha$, we
found that these flows develop normal stress differences for $\alpha >1$.

The value of the interparticle friction has an important role in the rheological
behavior of dense granular flows.  One example is that numerical simulations
with circular or spherical grains have shown\cite{cruz-pre-2005} that the effective 
friction in the low shear rate limit $\mu_c$ is an increasing function of $\mu_p$. 
Another one concerns the exponent of the constitutive law (see
Eq.~\ref{eq:muifit} below).  In the simulations of Favier et
al.\cite{favier-prf-2017} it is found that the exponent of the power law
term switches from $\beta=0.5$ in a
low friction limit ($\mu_p\lesssim 10^{-2}$) to
$\beta=1$ in the high friction limit ($\mu_p\gtrsim 10^{-1}$). Three regimes
have been identified and associated with different
dissipation mechanisms\cite{degiuli-pre-2016, trulsson-pre-2017}.

In this work we extend the $\mu(I)$ rheology to the case of a 3-dimensional 
system of frictional spherocylinders. We show, that in a realistic 3D 
frictional system the peculiar non-monotonic behaviour of $\mu(\alpha)$ is observed
in a much wider range of interparticle friction (up to around $\mu_p=0.4$, thus 
including common granular materials) than in 
the previously reported case of a simplified 2D system (up to $\mu_p=0.15$) 
\cite{trulsson-jfluidmech-2018}. We reveal the microscopic origin of this 
observation and relate the behaviour of the constitutive coefficients to the
above mentioned dissipation regimes. The paper is organised as
follows: in Section~\ref{sec:setup} we briefly recall the numerical setup of the
simulations, which builds on that of our earlier work\cite{nagy-pre-2017}.
In Section~\ref{sec:results} we present and discuss our numerical results, 
explain a number of observed phenomena, including scaling arguments to understand 
and interpret the data. We summarize and draw perspectives in 
Section~\ref{sec:summary}.

\section{Setup}
\label{sec:setup}

\begin{figure}[t]
  \centering
  \includegraphics[width=0.75\columnwidth]{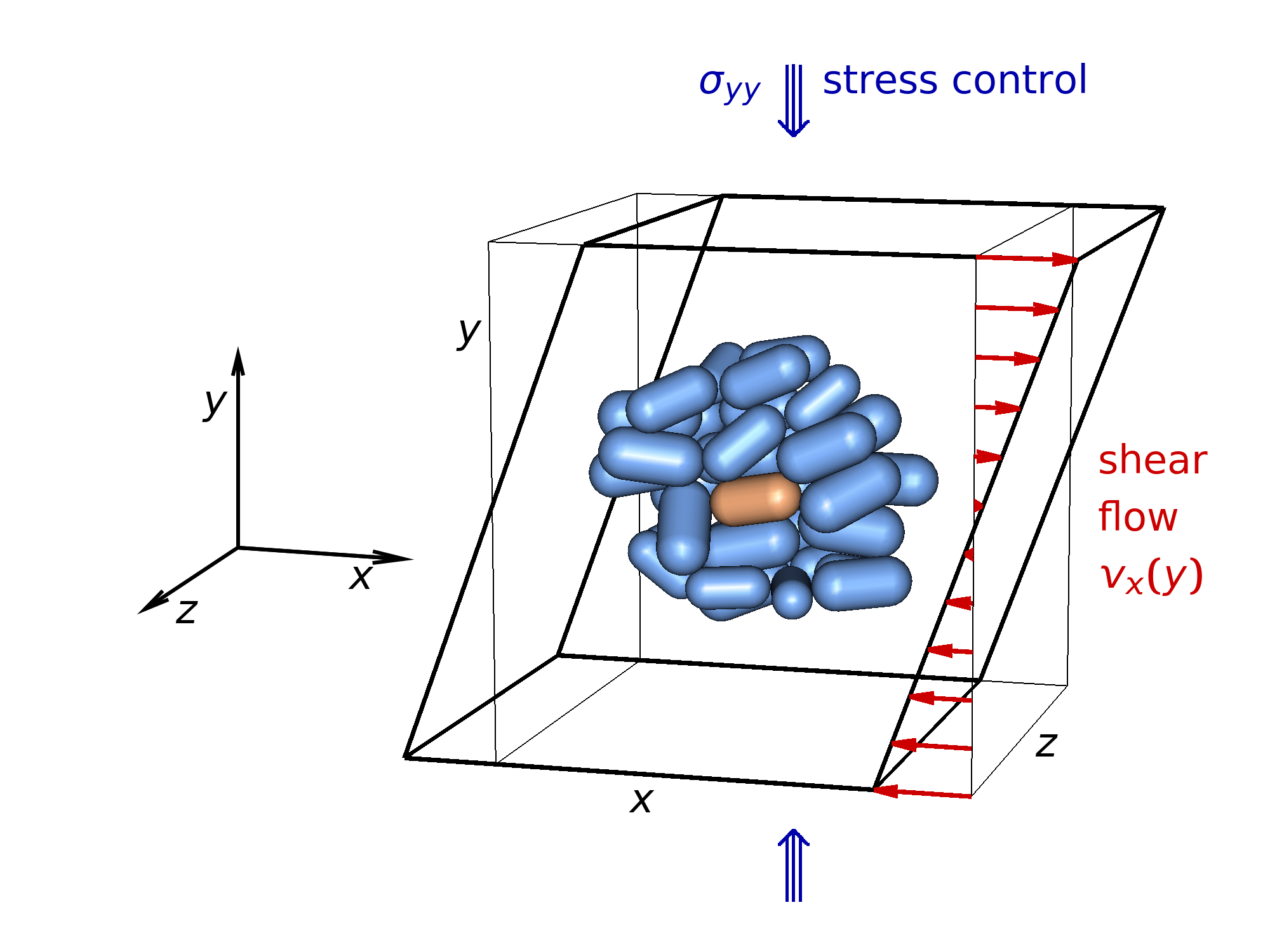}  %
  \caption{
    Setup of the simulation. Here we plot only one particle near the center of
    the box (orange) and particles in its hemispherical neighborhood (blue);
    their aspect ratio is $\alpha=2$.
    The system is sheared with shear rate $\dot\gamma=dv_x/dy$, and a
    constant normal stress $\sigma_{yy}=p_y$ is maintained by adjusting box side
    $L_y$ by a feedback loop.  
  }
  \label{fig:setup}
\end{figure}

We performed numerical simulations of homogenous shear flow in 3D plane
Couette geometry, see Fig.~\ref{fig:setup}.  
The particles were frictional spherocylinders with
length-to-diameter aspect ratio $\alpha = \ell / 2R$. The interparticle force
${\mathbf F}_{ij}$ which particle $j$ exerts on particle $i$ consists of a
soft core repulsion and a tangential component due to friction.  The repulsive
force points in the direction of the local surface normal $\hat c_{ij}$, its
amplitude is proportional
to the virtual overlap $\delta_{ij}$ between the particles, and it contains a
dissipative term proportional to the velocity difference ${\mathbf v}_{c,ij}$
at the contact point:
\(
{\mathbf F}_{ij}^\text{rep} = (-k\, \delta_{ij}
      + b\, {\mathbf v}_{c,ij}\cdot\hat c_{ij})\, \hat c_{ij} \,.
\)
The prefactor $b$ was determined by requiring a given coefficient of
restitution $e$ for binary collisions (we used $e=0.5$ in this paper).
The frictional tangential force is based on the Mindlin force law: the force
increment between time steps is 
\(
\Delta {\mathbf F}_{ij}^\text{fric} = k \Delta {\mathbf t}_{c,ij} \,,
\)
where $\Delta {\mathbf t}_{c,ij}$ is the tangential displacement (projected to
the plane perpendicular to $\hat c_{ij}$) during a time step between the
touching contact points of the particles.  The magnitude of the frictional
force is limited by the interparticle friction coefficient: 
\(
|{\mathbf F}_{ij}^\text{fric}| < \mu_p |{\mathbf F}_{ij}^\text{rep}|\,.
\)

The length, time and mass units of the simulation were set implicitly by
setting the mean particle diameter $2R$, density $\rho$ and contact stiffness
$k$ (equal for the normal and tangential force) to unity.  To prevent
crystallization for frictionless particles \cite{somfai-epjwebconf-2017},
especially at larger values of $\alpha$, we used size polydispersity of 10\%
(standard deviation to mean ratio in a uniform distribution). While
crystallization was less critical for frictional particles, we kept the
polydispersity fixed for consistency.  The rheological measurements were
performed under fixed normal stress: the $p_y := -\sigma_{yy}$ component of the
stress tensor (where $y$ is the velocity gradient direction) was controlled
around a fixed value of $10^{-3}$ by a feedback loop adjusting the $L_y$ side
of the periodic simulation box.

The rest of the simulation details, including the preparation protocol for
the initial conditions, are detailed in Ref.~\citenum{nagy-pre-2017}.

\section{Results}
\label{sec:results}

\subsection{Rheology}

\begin{figure*}[t]
  \includegraphics[width=\textwidth]{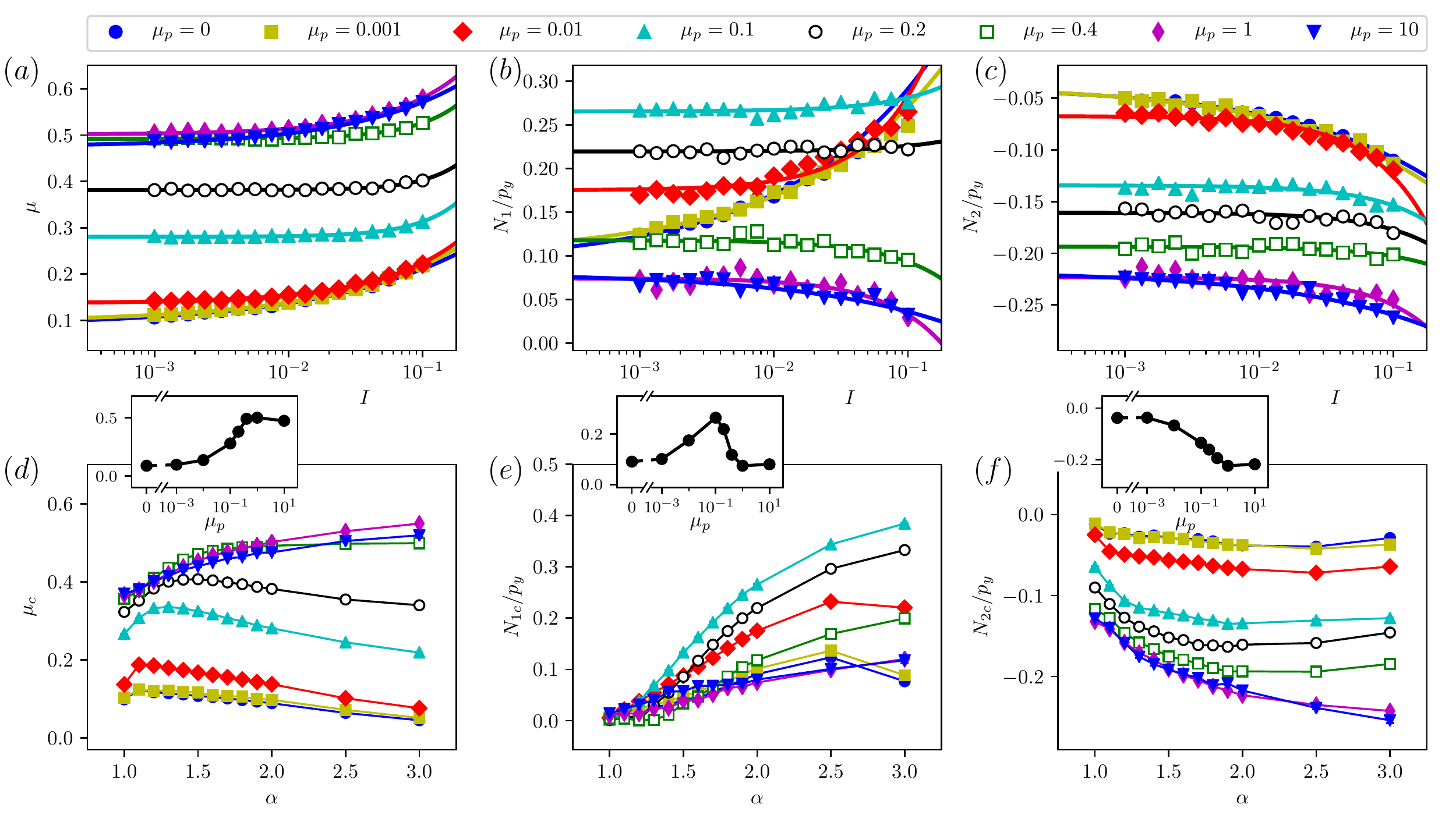}  %
  \caption{
        The inertial number dependence of rheological parameters are shown on
        top row for $\alpha=2$: 
        (a) effective friction,
        (b) normalized first normal stress difference, and
        (c) normalized second normal stress difference.
        The colors and symbol shapes indicate different interparticle friction
        coefficient values, including frictionless ($\mu_p=0$) and frictional
        in the range $10^{-3} \le \mu_p \le 10$.
        The quasistatic limit ($I\to 0$) of the same quantities are plotted in
        the bottom row for a range of shapes $1\le \alpha \le 3$; on the insets
        also the quasistatic limit values are plotted, but for $\alpha=2$ and
        against $\mu_p$, see text for details.
  }
  \label{fig:stress}
\end{figure*}

\begin{figure*}[!t]
  \includegraphics[width=\textwidth]{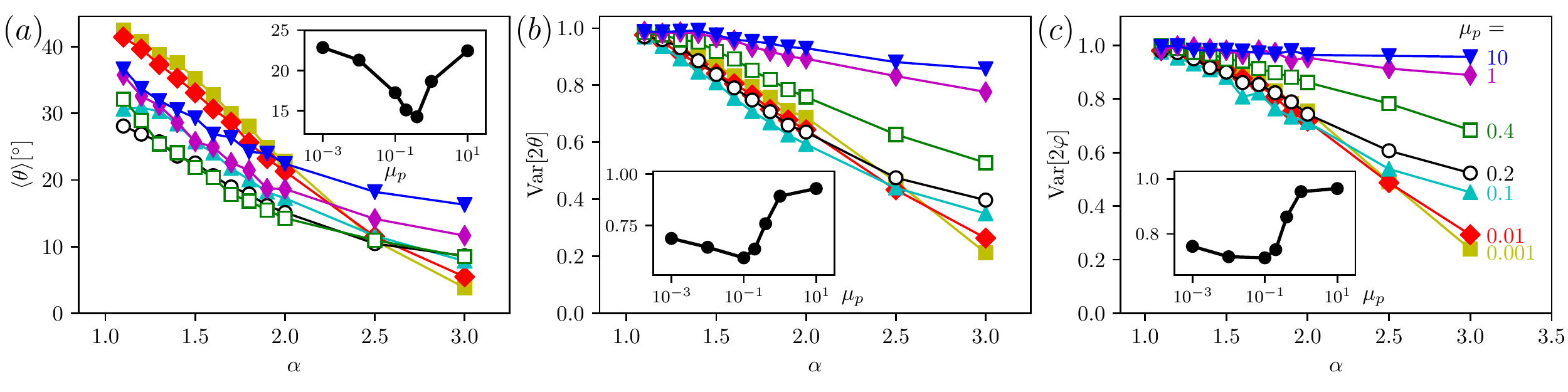}  %
  \caption{
      The aspect ratio dependence of (a) the average of the shear alignment
      angle $\theta$, (b) circular variance of $2\theta$, and (c) circular
      variance of $2\varphi$.  The angle $\theta$ denotes the deviation angle
      of the particle orientation from the streaming direction towards the
      gradient direction, and $\varphi$ is the deviation towards the neutral
      direction.  Circular variance of $1$ corresponds to completely uniform
      angular distribution, while $0$ corresponds to zero-width (Dirac-delta)
      distribution.  Symbols and colors represent different $\mu_p$.
      The insets, like on Fig.~\ref{fig:stress}, show the dependence on
      $\mu_p$ for aspect ratio value $\alpha=2$.
  }
  \label{fig:angle}
\end{figure*}

The inertial number dependence of the following rheological parameters are
shown on the top row of Fig.~\ref{fig:stress}: effective friction
$\mu=\sigma_{xy}/p_y$, normalized first normal stress difference $N_1/p_y =
(\sigma_{xx}-\sigma_{yy})/p_y$, and normalized second normal stress difference
$N_2/p_y = (\sigma_{yy}-\sigma_{zz})/p_y$. The solid curves are fit (in
the range $10^{-3}\le I \le 10^{-1}$) to the following empirical form:
\begin{equation}
  \mu(I) \approx \mu_c + \mu_1 I^\beta \;,
  \label{eq:muifit}
\end{equation}
and similarly for $N_1/p_y$ and $N_2/p_y$. The values of the exponent $\beta$
ranged between about 0.4 and slightly more than 1, where we observed the
smallest values for the frictionless case, and the largest values for
moderate friction $0.1 \lesssim \mu_p \lesssim 0.4$.
Equation~(\ref{eq:muifit}) enables the reliable extraction of the quasistatic
($I\to 0$) limits, which are plotted in the bottom row of
Fig.~\ref{fig:stress}.  

\subsubsection{Effective friction}

One remarkable finding is that the quasistatic
effective friction coefficient $\mu_c$ is a non-monotonic function of the
aspect ratio for negligible to moderate interparticle friction ($\mu_p
\lesssim 0.4$), and the aspect ratio where the maximum occurs shifts to larger
values for increasing $\mu_p$.  For large $\mu_p$ the effective friction is a
monotonically increasing function of the aspect ratio, at least in the range
$1 \le \alpha \le 3$ we explored.%
\footnote{
We note that $\mu_c$ extrapolated from high inertial number non-homogenous
flow down an inclined plane\cite{hidalgo-prf-2018} (for $\mu_p=0.5$) is
remarkably close to our measurements. 
}

As discussed above, the nonmonotonicity of $\mu_c(\alpha)$ has been observed 
earlier for purely frictionless spherocylinders \cite{nagy-pre-2017}, and for 
2D ellipses with low friction coefficient \cite{trulsson-jfluidmech-2018}. 
The data in Fig.~\ref{fig:stress}(d) clearly show, that in a 3D system this is 
observed in an extended friction range, which already includes realistic materials.  We now explore the microscopic ingredients leading to this behaviour.

The explanation can be traced back to the shear induced alignment of elongated
particles, initially described in Refs.~\citenum{borzsonyi-prl-2012} and
\citenum{borzsonyi-pre-2012}.
Let us denote the deviation of the particle axis from the streamlines by
$\theta$ within the $x$-$y$ plane, and by $\varphi$ out of this plane.
Due to shear the elongated particles develop nematic ordering, where
$\left<\theta\right>$, which we call shear alignment angle, is interestingly
non-zero [see Fig.~\ref{fig:angle}(a)]
\footnote{\NEW{The periodicity of $\theta$ by $\pi$ must be taken into account when
calculating its average. This can be done by calculating the nematic order
tensor $\langle(3/2)\hat e\circ \hat e - 1/2\rangle$ (where $\hat e$
is the unit vector in the particle's axis) and considering its largest
eigenvalue's eigendirection, which we do to obtain Fig.~\ref{fig:angle}(a), or
by averaging on the complex unit disk:
$(1/2)\arg\langle\exp(2i\theta)\rangle$.}}%
, while $\left<\varphi\right>=0$ by
symmetry.  (With increasing elongation $\alpha$ the distributions of these
angles become typically narrower, see Fig.~\ref{fig:angle}(b-c). (Due to the
periodicity directional statistics have to be used, and since both $\theta$
and $\varphi$ are periodic by $\pi$, the relevant quantities are the circular
variances $\text{Var}[2\theta]$ and $\text{Var}[2\varphi]$.)
For small $\mu_p$ the circular variances drop sharply with $\alpha$, resulting
in more orientationally ordered configurations.  In addition
$\left<\theta\right>$ decreases as well, which altogether leads to a situation
where the particles obstruct each other's motion less, thus despite the more
elongated shape the shear resistance $\mu_c$ decreases.  For larger particle
friction however, $\text{Var}[2\theta]$ barely decreases with $\alpha$, and
the drop in $\text{Var}[2\varphi]$ is also very small; these packings remain
orientationally rather disordered. The disoriented particles with increasing
elongation hinder each other's motion more, leading to a monotonically
increasing $\mu_c$ as a function of $\alpha$.

\begin{figure}[!t]
  \begin{center}
  \includegraphics[width=0.6\columnwidth]{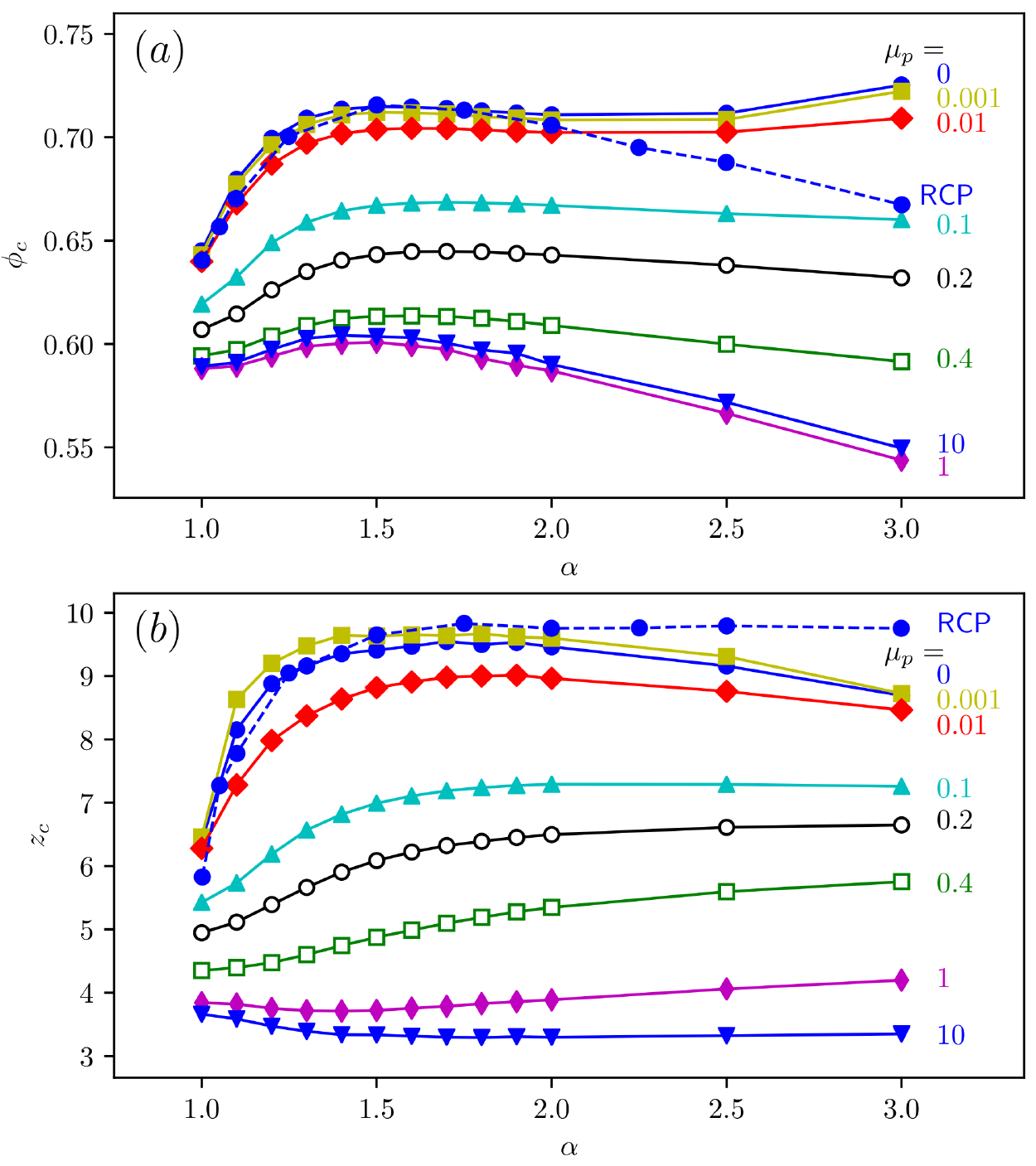}  %
  \end{center}
  \caption{
        The quasistatic value of the (a) volume fraction $\phi_c$ and
        (b) coordination number $z_c$, plotted against the aspect ratio
        $\alpha$ for different interparticle friction coefficient values.
        The symbols  are the same as in Fig.~\ref{fig:stress}, with
        the blue full circles with dashed lines
        corresponding to random close packed simulations without particle
        friction in non-sheared systems \cite{donev-science-2004}.
  }
  \label{fig:phiz}
\end{figure}

\subsubsection{Normal stress differences}

The first normal stress difference
[Fig.~\ref{fig:stress}(e)] is zero for spherical particles regardless of
friction, and increases monotonically with aspect ratio (except for very small
$\mu_p$).  The second normal stress difference [Fig.~\ref{fig:stress}(f)] is
nonzero even for spherical particles.

The insets of the bottom row of Fig.~\ref{fig:stress} present the
effective friction and the normal stress differences as a function of the
interparticle friction coefficient $\mu_p$ for $\alpha=2$. As it is expected
the effective friction of the system is first gradually growing with
increasing $\mu_p$, but we find an interesting unexpected breakdown of this
growing trend between $0.4<\mu_p<1$.  We also see, that $N_1$, i.e. the normal
stress
in the gradient direction (with respect to the flow direction) reaches a peak
at $\mu_p=0.1$, where $\mu_c$ has the highest growth rate, and then decreases
back to a small value. At the same time $N_2$, the normal stress in the neutral
direction with respect to the gradient direction also reaches its minimum.
This can be explained by considering the strong change in the particle
orientational order, and the distribution of the forces on the particles.

The first and second normal stress differences can be rewritten in terms of
average particle level quantities, like orientational angles, their
variance, and quantities related to the force distribution on the particle
surface.  
\NEW{
In order to obtain an analytical expression to the stress differences, we
have to make approximations, most importantly neglect correlations
between the in plane angle $\theta$, the out of plane angle $\phi$, and the
eigenvalues of the single particle stress tensor.
By doing so we get curves that closely resemble the values obtained by the
simulation, including their dependence on $\alpha$ and $\mu_p$ in most cases
} 
[Fig.~\ref{fig:stress}(e,f) and their insets].
In particular, these calculations
recover that $N_1=0$ for $\alpha=1$ (with the exception of very small
$\mu_p$); that $N_1$ is increasing and $N_2$ is decreasing function of $\alpha$,
the dependence of $N_1$ on $\mu_p$ is non-monotonic (for example for
$\alpha=2$), and the decreasing trend of $N_2$ on small to medium values of
$\mu_p$.
\NEW{
These derivations are technical and we have gathered them in the Supplementary
Material accompanying this article.
}

\subsubsection{Volume fraction and coordination number}

To complete the rheological description, the quasistatic values of the volume
fraction $\phi_c$ and the coordination number $z_c$ are plotted as
a function of the aspect
ratio on Fig.~\ref{fig:phiz}.  It is interesting to note, that for the packing 
fraction the random close packed (RCP) values, obtained as simulation of
frictionless particles without shear \cite{donev-science-2004}, follow closely
our $\mu_p=0$ case for $\alpha \lesssim 2$, but deviate for larger aspect
ratios: our values start to increase while RCP shows decreasing trend for
growing $\alpha$.  For moderate interparticle friction the packing fraction
of sheared spherocylinders
also decreases for large $\alpha$, but the curves are still shallow. Our
explanation is the following. The most significant difference between our
measurements and RCP is that in our case the system is sheared, while for RCP
it is not.  Shear induces orientational order for elongated particles
\cite{borzsonyi-prl-2012}, which gets increasingly pronounced for larger
aspect ratios, and orientational order increases the packing fraction.  Similar
effect (sheared spherocylinders and RCP deviates only for $\alpha \gtrsim
2$) is observed for the coordination number $z_c$ as well.
\NEW{
A volume controlled simulation has been performed recently by Nath and
coworkers \cite{nath-epje-2019}; their jamming density agrees with our
measurements using stress controll, which shows the robustness of these
results.}

\subsection{Dissipation regimes}

\begin{figure*}[h]
  \begin{center}
    \includegraphics[width=\textwidth]{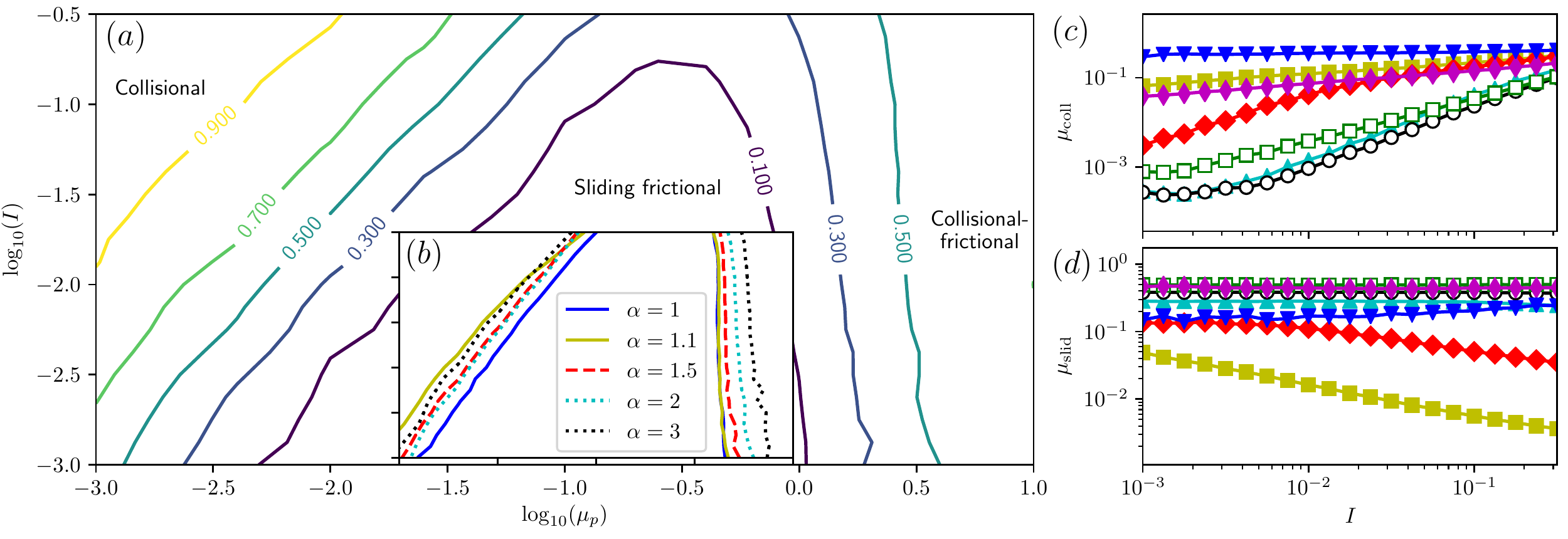}  %
    \vspace*{-8mm}
  \end{center}
  \caption{
        (a) The fraction of the collisional loss of the total dissipation as a
        function of the inertial number and interparticle friction coefficient,
        plotted as level sets.  
        (b) Dependence of the 50\% level set on the aspect ratio in the range
        $1\le\alpha\le 3$.
        (c) Collisional and (d) sliding friction contribution to the effective
        friction coefficient as a function of the inertial number for a range
        of interparticle friction coefficient values.  The power law exponent
        depends on the dissipation regime.
        The aspect ratio on panels (a), (c) and (d) is $\alpha=2$.
        The symbols on panels (c) and (d) are the same as on
        Fig.~\ref{fig:stress}.
  }
  \label{fig:regime}
\end{figure*}

\begin{figure}[tb]
  \centering
  \includegraphics[width=0.6\columnwidth]{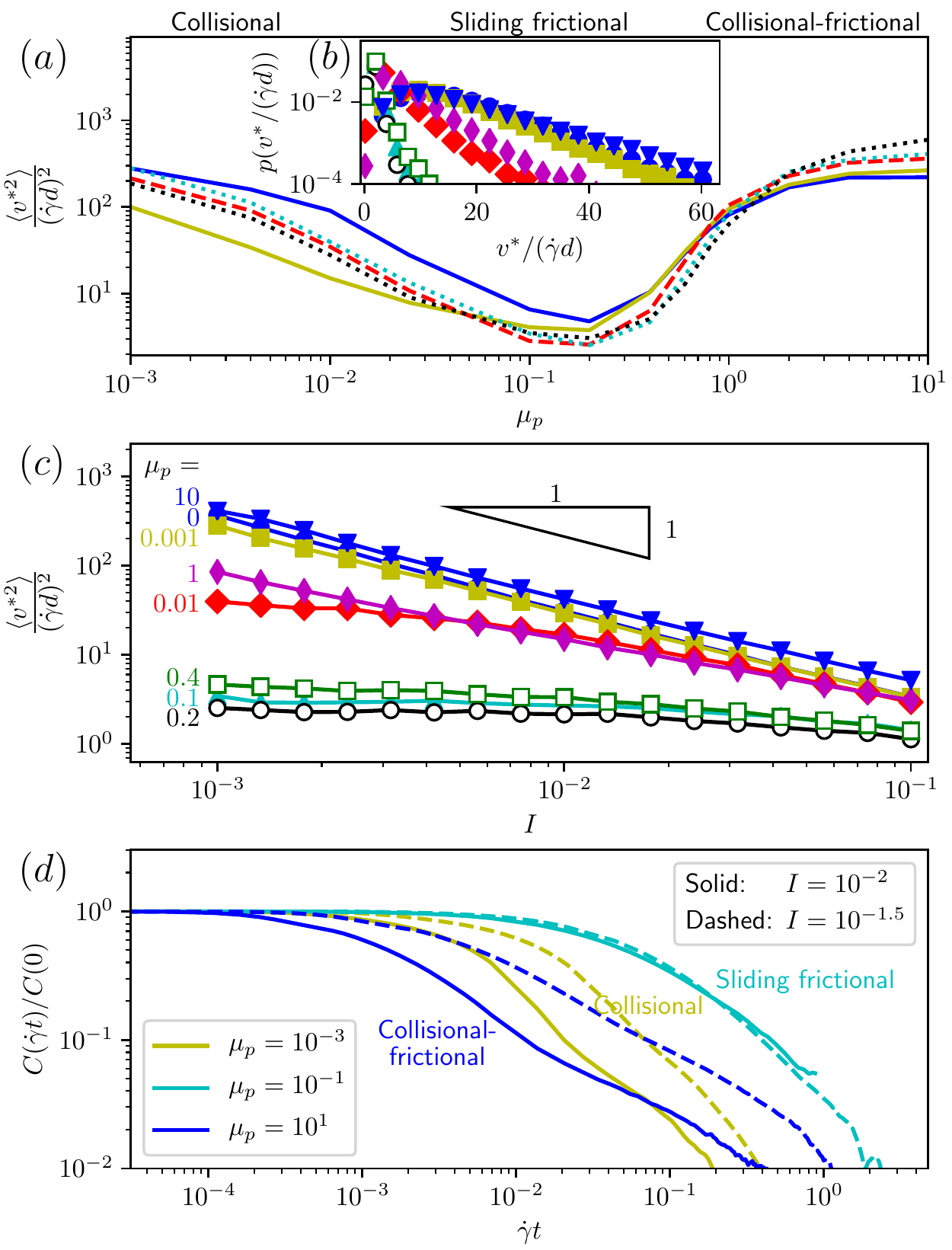}  %
  \vspace*{-5mm}
  \caption{
    (a) Variance of the random velocity $v^*$, non-dimensionalized by the shear
	rate, as a function of $\mu_p$, for $I=10^{-3}$. Line colors represent
	different $\alpha$ values, as in the legend of Fig.~\ref{fig:regime}(b). (b)
	Distribution of $v^*$ for $\alpha=2$. (c) Variance of the random velocity
	$v^*$, as a function of the inertial number. The symbols on panels (b) and
	(c) represent the same values of $\mu_p$ as in
        Figs.~\ref{fig:stress}-\ref{fig:phiz}.
        (d) Normalized autocorrelation of the random velocity for $\mu_p$
        values in the three regimes (colours) and two different inertial
        number values (continuous and dashed lines).  
        \NEW{The autocorrelation function in the sliding friction regime
        ($\mu_p=10^{-1}$ curve) fall off by
        $\dot\gamma t$, so the curves for different $I$ collapse. However, in
        the collisional and the collisional-frictional regimes
        ($\mu_p=10^{-3}$ and $10^1$) the dependence is on $\dot\gamma t/I$;
        the curves would approximately collapse when shifted by the difference
        in $I$.
        }
  }
  \label{fig:velocity}
\end{figure}

The flow of granular materials is -- like any typical material -- dissipative,
quantified by the effective friction coefficient $\mu$.  For granular
materials dissipation has two sources: collisional loss (parameterized by the
coefficient of restitution $e$) and sliding friction (parameterized by the
interparticle friction coefficient $\mu_p$).  Since the two mechanisms have
different nature, it is worth considering which one is dominant as a function
of the parameters \cite{degiuli-pre-2016}.  When comparing the two mechanisms,
we keep the coefficient of restitution at a fixed intermediate value $e=0.5$,
and vary $\mu_p$ and $I$.

\subsubsection{Regime diagram}

Figure~\ref{fig:regime}(a) shows the fraction of the collisional loss 
\NEW{(calculated on the contacts during force evaluation)} to the
total loss as a function of $I$ and $\mu_p$.  The 50\% level set divides the
parameter space into three regimes:
\emph{collisional} loss \NEW{dominated} for large $I$ and small $\mu_p$,
\emph{sliding friction} dominated for intermediate values of $\mu_p$, and
a third regime, which we call \emph{collisional-frictional}
\footnote{While Ref~\citenum{degiuli-pre-2016} calls this region ``rolling'',
we consider \emph{collisional-frictional} more appropriate, as the dominant
contribution to dissipation is collisional, and not rolling dissipation.},
where sliding friction is somewhat suppressed as $\mu_p$ is so large that the
contacts are rarely sliding.  Since the transitions are smooth, we call these
regions as \emph{regimes}, instead of phases (which would imply sharp
transition).  
These regimes have been identified earlier for spherical particles
\cite{degiuli-pre-2016,trulsson-pre-2017}; here we confirm their presence for
elongated grains, and investigate how they are affected by changing the
particle elongation.
Figure~\ref{fig:regime}(b) shows the dependence of the 50\% level set on the
aspect ratio.  The borderline between sliding and collisional-frictional
shifts to larger $\mu_p$ for more elongated particles: the nearly spherical
particles stop sliding at around $\mu_p\approx 2$, while for $\alpha=3$
this only happens beyond $\mu_p\approx 4$ or $5$.  The elongated particles are
more entangled by their neighbors, their rotational degrees of freedom are
suppressed, resulting in a decrease in collisional dissipation, so sliding
friction dominated regime extends further.
The border between collisional and sliding friction dominated regime also
shows slight aspect ratio dependence, but there it is not monotonic on
$\alpha$.

Figure~\ref{fig:regime}(c) and (d) show the dependence of the collisional and
the sliding friction contribution to the effective friction as a function of
the inertial number.  Power law scaling can be observed, with the scaling
exponents depending on the dissipation regime.  This is especially striking for
$\mu_p=0.01$, where increasing $I$ switches regimes from sliding to
collisional at around $I=10^{-2}$.

\subsubsection{Velocity fluctuations}
\label{sec:velocityfluctuation}

The velocity fluctuations also depend strongly on the dissipation regime.
On Fig.~\ref{fig:velocity}(b) the distribution of the random velocity is
shown ($v^*$ is the excess velocity on top of the linear velocity profile).
The velocity distribution displays a heavy tail in both the collisional and
the collisional-frictional regimes. The same phenomenon is displayed by the
width (second moment) of the velocity distribution
(Fig.~\ref{fig:velocity}(a)). 
Not only the value, but also the $I$-dependence of the velocity
fluctuations varies by the dissipation regime.  Figure~\ref{fig:velocity}(c)
shows that $\big< {v^*}^2 \big> /(\dot\gamma d)^2 \sim I^{-1}$ in the
collisional and collisional-frictional regimes, while practically independent
of $I$ (i.e., $\sim I^0$) in the sliding friction regime.  Below we
provide scaling arguments based on the underlying microscopic processes, which
also explain the behavior of the autocorrelation functions plotted on
Fig.~\ref{fig:velocity}(d).

The velocity fluctuations can be understood by the following simple 
microscopic picture. One must distinguish between the regime of low and
large values of $\mu_p$ on the one hand, for which these fluctuations are
large, and the regime for intermediate values of $\mu_p$ on the other
hand, for which they are significantly smaller [Fig.~\ref{fig:velocity}(a)].
In the first
highly fluctuating case, the grains move in an intermittent way. They
experience short phases of typical duration $T = d/\sqrt{p/\rho}$ during
which they are suddenly accelerated by the pressure $p$ to a velocity
$d/T$ with respect to their neighbours. The average fluctuating kinetic
energy per grain then scales as $\rho d^3 (d/T)^2 \times f$, where $f$ is
the fraction of time during which this acceleration phase occurs, i.e. $f
= T\dot\gamma = I$. This argument \cite{gdrmidi-eurphysje-2004,
favier-prf-2017} gives $m\big< {v^*}^2 \big> \sim d^3 p I$.
Dividing by the average relative velocity, we obtain $\big< {v^*}^2
\big> /(\dot\gamma d)^2 \sim I^{-1}$, as shown in
Fig.~\ref{fig:velocity}(c).

By contrast in the second case, the particles' motion does not appear
intermittent. The grains move continuously ($f=1$), at a time scale that
follows the overall shear rate: $T \sim 1/\dot\gamma$. The average
fluctuating kinetic energy per grain then scales as $d^3 p I^2$, leading
to $\big< {v^*}^2 \big>/(\dot\gamma d)^2 \sim I^0$. This behaviour is
also consistent with the corresponding curves in Fig.~\ref{fig:velocity}(c),
which are almost flat for intermediate $\mu_p$.

This change in the relevant time scale $T$ is supported by the computation
of the autocorrelation function, displayed in Fig.~\ref{fig:velocity}(d).
For $\mu_p=10^{-1}$,
the curves lie above those for $\mu_p=10^{-3}$ and $\mu_p=10^1$,
indicating more persistent grain motion. Also, plotted as functions of
$\dot\gamma t$ the curves for intermediate $\mu_p$ show a collapse when
varying $I$, while the others rather follow the scale $t\sqrt{p/\rho}/d =
\dot\gamma t/I$.

\subsubsection{Fluctuations of local shear and rotation}

\begin{figure}[t]
  \centering
  \includegraphics[width=0.6\columnwidth]{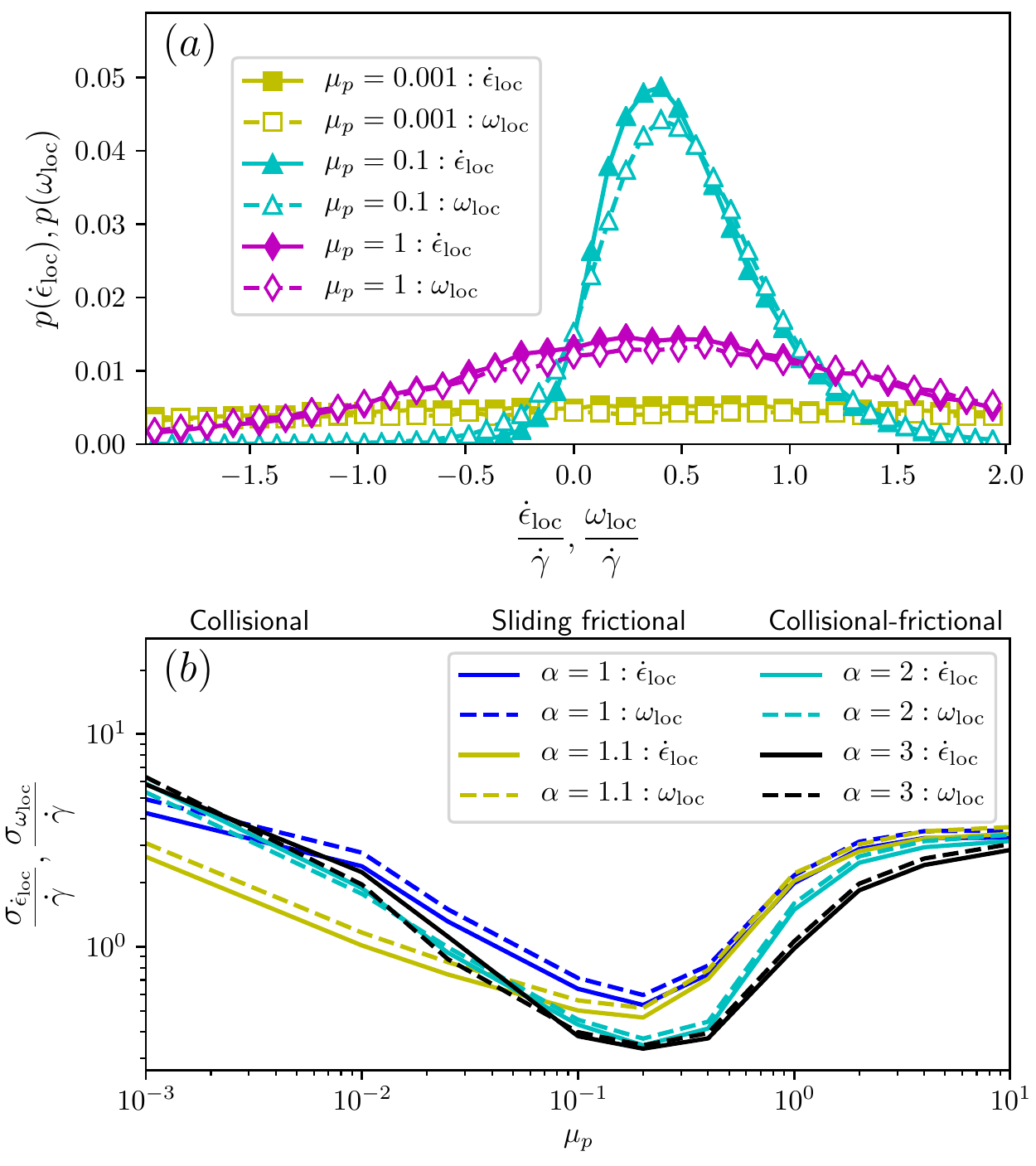}  %
  \caption{
        (a) Distribution of the local pure shear rate
        $\dot\epsilon_\text{loc}$ (solid symbols) and local
        angular velocity $\omega_\text{loc}$ (open symbols) on mesoscopic
        scales.
        (b) The distribution width (standard deviation) of the mesoscopic
        pure shear rate and angular velocity as a function of $\mu_p$.
        Both quantities have relatively narrow distribution in the sliding
        frictional regime, and have very wide distribution in the other two
        regimes.
  }
  \label{fig:deform}
\end{figure}

The fluctuations display similar trend on the mesoscopic scale as well.
To extract local deformation rates, the simple shear can be written as a
sum of pure shear and solid body rotation:
\[
  \dot{\pmb\gamma} = 
  \left(\begin{array}{ccc}
    0 & \dot\gamma & 0 \\
    0 & 0 & 0 \\
    0 & 0 & 0 \\
  \end{array}\right)
  =
  \left(\begin{array}{ccc}
    0 & \dot\epsilon & 0 \\
    \dot\epsilon & 0 & 0 \\
    0 & 0 & 0 \\
  \end{array}\right)
  +
  \left(\begin{array}{ccc}
    0 & \omega & 0 \\
    -\omega & 0 & 0 \\
    0 & 0 & 0 \\
  \end{array}\right) \;,
\]
so $\dot\epsilon_\text{loc}$ and $\omega_\text{loc}$ can be obtained as the
symmetric and antisymmetric $xy$ component of the local deformation rate
tensor. For homogenous simple shear, $\dot\epsilon=\omega=\dot\gamma/2$.
Figure~\ref{fig:deform}(a) shows the distribution of the local pure shear
rate and the local angular velocity of mesoscopic regions. \NEW{The local
strain rate tensor is obtained by linear regression of the relevant components
of the matrix, which projects the particle positions onto their velocity
space. The particles are sampled from localized regions}
of linear extent $1/4$ of the largest side of the simulation box. As
evident from this panel and Fig.~\ref{fig:deform}(b) (the distribution width
of the same quantities), both the local pure shear rate and the local angular
velocity have moderately narrow distribution around their mean (which is $1/2$
for both the normalized $\dot\epsilon_\text{loc}/\dot\gamma$ and
$\omega_\text{loc}/\dot\gamma$) in the sliding frictional regime, while the
distribution is very wide in the other two regimes. This includes
non-negligible fraction of mesoscopic regions, which deform and/or rotate with
opposite sign compared to the bulk average.

Similarly to the velocity fluctuations,
due to the intermittency of grain motion at small and large $\mu_p$,
the width of these distributions around their average values are an order of
magnitude larger than that for intermediate interparticle friction. In the
latter case, the grain's velocity is then typically affine, following the
global shearing dynamics.

\begin{figure*}[t]
  \includegraphics[width=0.48\textwidth]{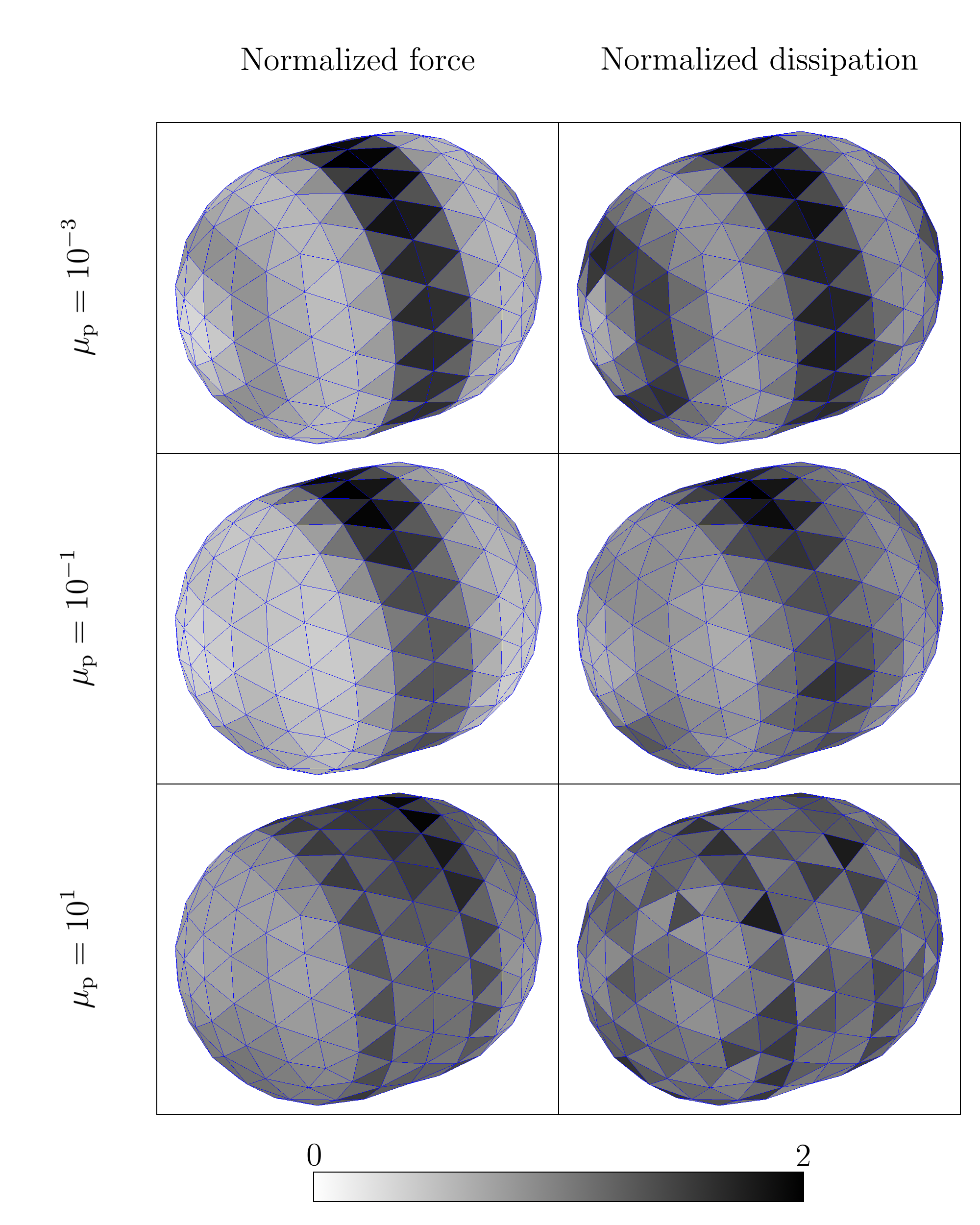}  %
  \includegraphics[width=0.48\textwidth]{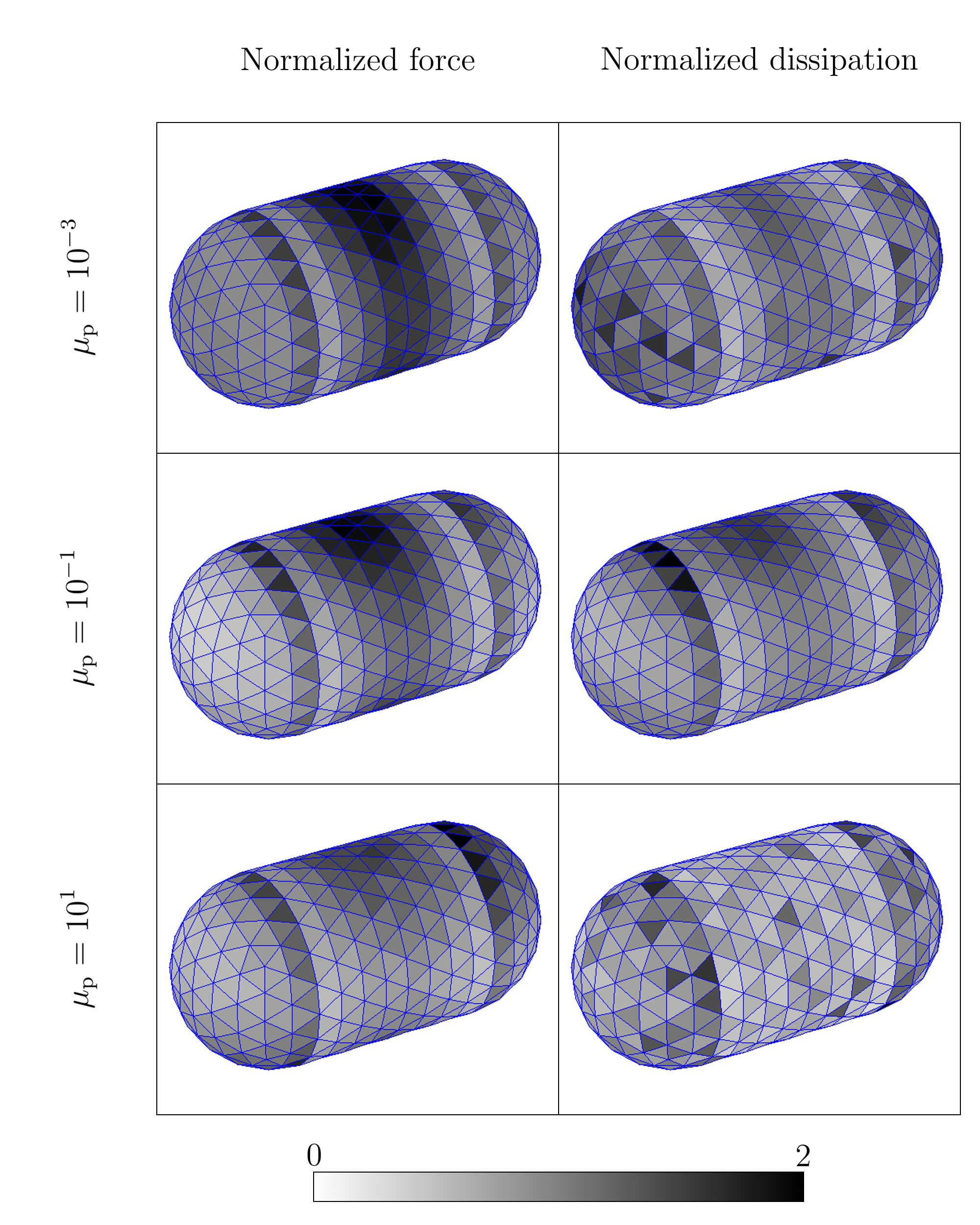}  %
  \caption{
      
        The spatial distribution of forces and the dissipation on the surface
        of the particles. On the left panel the particles are only slightly
        elongated, aspect ratio is $\alpha=1.3$; this is the shape
        where the effective friction $\mu_c$ is maximal for $\mu_p=0.1$.
        The right panel displays more elongated particles with $\alpha=2$,
        corresponding to the shape used on the insets of Fig.~\ref{fig:stress}.
        On both panels the first columns show the force distribution on the
        surface of a particle,
        non-dimensionalized as $|\left\langle{\mathbf F}_k\right\rangle| / A_k
        \sigma_{yy}$, where the numerator is the absolute value of the
        vectorial average of the forces on the $k$-th surface element, and $A_k$
        is the area of the surface element. The second columns show the
        dissipation
        distribution non-dimensionalized as $N\langle P_k\rangle A /
        \sigma_{xy} \dot\gamma V_{\text{box}} A_k$, where $P_k$ is the power
        dissipated on the surface element, $N$ is the number of particles, $A$
        is the surface area and $V_{\text{box}}$ is the volume of the box. For
        this figure $\alpha=1.3$ and $I=10^{-3}$.
        The normalization in the first columns is fixed across the three
        images ($\sigma_{yy}=10^{-3}$), while in the second columns it is
        different ($\sigma_{xy}$, like $\mu_c$, varies by a factor of 
        $\approx 4$ across small to large $\mu_p$).
  }
  \label{fig:mesh}
\end{figure*}

\subsubsection{Force and dissipation spatial distribution}

It is interesting to consider which part of the surface of the particles
experience the strongest confining forces, and whether these areas coincide
where most of the dissipation takes place.  In the left panel's first column of
Fig.~\ref{fig:mesh} the normalized forces on an $\alpha=1.3$ particle are
displayed: the absolute value of the vectorial average of the
forces acting on a surface element is normalized by a typical force based on
the confining pressure. In the frictionless regime the forces are concentrated
on the cylindrical belt, which holds, somewhat less sharply, in the sliding
friction regime as well.

This concentration of the forces on the cylindrical belt 
can be understood by looking at the torque. For
spherocyliders the torque on particle $i$ can be expressed as:
\begin{equation}
\mathbf{\tau}_i = \sum_{j \in Z^{(i)}} \left( \lambda_{ij} \mathbf{o}_i  + d (1 
- \frac{1}{2} \delta_{ij}) \hat{c}_{ij} \right) \times | 
\mathbf{F}_{ij}^{\text{rep}} | \left( \hat{c}_{ij} + \mu_{\text{p}} \zeta 
\hat{t}_{\text{c},ij} \right) \;,
\end{equation}
where $\mathbf{o}_i$ is the orientation vector of the particle, $Z^{(i)}$ is
the set of particles that particle $i$ is in contact with, $\lambda_{ij} \in
\left[- d \left(\alpha - 1 \right),  d \left(\alpha - 1 \right) \right]$ is
the (signed) distance between the centre of mass and the normal projection of
the contact point to the symmetry axis, and $\zeta 
= |{\mathbf F}_{ij}^\text{fric}| / \mu_p |{\mathbf F}_{ij}^\text{rep}|
\in \left[0, 1\right]$ is the friction mobilization or plasticity index. In
the frictionless case ($\mu_{\text{p}} = 0$) only one component remains (since
$\hat{c}_{ij} \times \hat{c}_{ij} = 0$):
\begin{equation}
\mathbf{\tau}_i = \sum_{j \in Z^{(i)}} \lambda_{ij} | 
\mathbf{F}_{ij}^{\text{rep}} | \mathbf{o}_i \times \hat{c}_{ij} \;.
\end{equation}
Based on this expression the torque of a single contact is zero only at the two 
singular points at the tips of the particle $(\mathbf{o}_i \times \hat{c}_{ij} 
= 0)$ which are unstable, and at the circle along the centre of the cylinder 
$(\lambda_{ij} = 0)$ which is stable. This shows that frictionless particles,
\NEW{regardless their aspect ratio,}
prefer to align in a way that their contact points (especially those
carrying large forces) are at the middle of the cylindrical belt,
\NEW{as simultaneous force and torque balance is easier achieved when many of
the torque contributions are small; this is especially apparent when only a
few (eg. two) contacts dominate the force balance on a particle.}
This argument is valid with good approximation for small nonzero 
$\mu_{\text{p}}$, as the Coulomb cone is still restricted close to the normal 
direction ($\mu_{\text{p}} \zeta \ll 1$).
In the collisional-frictional regime, however, the areas experiencing the
largest forces are the edges of the spherical caps. 

The second column on the left panel of Fig.~\ref{fig:mesh} shows the distribution of the
dissipation, also normalized by the fraction of the total dissipation
projected onto a surface element. There is strong correlation between areas of
large forces and large dissipation, with two remarks. First, in the
collisional regime there is an extra high-dissipation ring on each of the
spherical caps, their location is determined by geometrical constraints on the
neighboring particles.  Second, the noise in the obtained distribution is
largest in the collisional-frictional regime. This can be explained by the
fact that the contacts are long lasting in this regime, so fewer separate
contacts contribute to the average; and unlike in the sliding friction regime,
the contacts are more often localized to their original formation point, and
thus contribute only to a single surface element.

On the right panel of Fig.~\ref{fig:mesh} we show the same quantities for a
more elongated, $\alpha=2$ particle shape. We can observe, like for less
elongated particles, that the forces are concentrated on the middle of the
cylindrical belt, especially for small to intermediate particle friction.
For
$\mu_p=10^{-3}$, the spherical caps receive also significant incoming forces,%
\footnote{\NEW{Similar increased concentration of contacts on the central
region at small aspect ratios, and on the tips at large aspect ratios are also
observed on frictionless 2D elongated particles\cite{marschall-pre-2018}.}}
while for $\mu_p=10^{-1}$ they do not.
This can be explained by considering that the elongated particles, regardless
of $\mu_p$, are oriented by shear to an average direction which is close to
the streaming ($x$) direction.  We have seen in
Section~\ref{sec:velocityfluctuation} that in the collisional regime the
velocity fluctuations are large; for oriented elongated particles this
fluctuation mainly happens in the axial direction of the particles%
\footnote{We confirmed that the velocity fluctuations are indeed larger in the
$x$ direction compared to the $y$ and $z$ directions.}, causing frequent
collisions on the caps.  For the sliding friction
regime the velocity fluctuations are much smaller, therefore the collisions on
the caps are less frequent and weaker, resulting in less incoming force on
those surface elements.  This in fact explains why
$N_1=\sigma_{xx}-\sigma_{yy}$ is small in the
collisional regime and large in the sliding friction regime: up to moderate
particle friction the Coulomb cone of the forces is narrow, the direction of
the contact forces is close to the surface normal, which for particles
oriented with their axis close to the $x$ direction means that forces on the
spherical caps contribute mostly to $\sigma_{xx}$, while those on the top and
bottom of the cylindrical belts contribute to $\sigma_{yy}$. In the
collisional regime the difference is small, while in the sliding friction
regime it is large as the caps receive little incoming force.  In the high
particle friction collisional-frictional regime both the Coulomb cone
is very wide and the orientational ordering is weak, thus
the forces can point in almost any direction, resulting in a more isotropic
force direction distribution, therefore again small $N_1$.  This completes the
explanation of the non-monotonic behavior of $N_1$ on $\mu_p$ for elongated
particles [Fig.~\ref{fig:stress}(e) inset].

\section{Summary and perspectives}
\label{sec:summary}

We performed DEM simulations to investigate the rheology of 
a realistic 3-dimensional frictional granular material consisting of
elongated particles (spherocylinders). Such systems develop orientational 
ordering when exposed to shear flow. The degree of this ordering depends 
on the interparticle friction and particle elongation on a nontrivial manner.
Namely, the shear induced orientational ordering is in principle increasing 
with particle elongation, but the characteristics of collisional and frictional 
interactions between neighbours (which hinder each others rotation) changes
with the interparticle friction coefficient.
We measured how key rheological quantities, including effective friction and 
normal stress differences depend on these two key parameters.
We found that the aspect
ratio dependence of the effective friction is non-monotonic not only for
frictionless particles as we saw earlier, but also for 
frictional particles up to $\mu_p\lesssim 0.4$, -- a range already
relevant for every day materials. For higher $\mu_p$ the effective friction is
monotonically increasing. We explained the microscopic origins of both the
non-monotonic behavior for small and intermediate $\mu_p$ and the monotonic 
one for large $\mu_p$. These observations are connected to the fact, that
for small friction coefficient the increasing particle aspect ratio
leads to stronger ordering and smaller average alignment angle --
consequently less obstruction between particles -- leading to less resistance
against shearing. For particles with large surface friction, however, for
increasing aspect ratio the stronger entanglement is not counteracted by the
ordering -- as it is weaker in this case -- leading to monotonically 
increasing shear resistance.
We showed that the collisional, sliding frictional, and
collisional-frictional dissipation regimes, which have been identified before
for spherical particles, are found also for elongated ones, and observed that
the boundary between the sliding frictional and the collisional-frictional
regimes moves towards higher $\mu_p$ for increasing aspect ratio, i.e. 
increasing grain elongation leads to the expansion of the sliding frictional
regime to higher values of the interpaticle friction. 
We explain this by considering the effect of entanglement on motion: for
more elongated particles the larger entanglement leads to the suppression of
the rotational motion, shifting the balance from collision-dominated to
sliding friction dominated dissipation.
We observed that the velocity fluctuations behave
differently in the dissipation regimes, and explained its microscopic origins
based on the different characteristic time scales of
the fluctuations.
We measured the spatial distribution of the forces on the
particle surfaces, and observed that for small $\mu_p$ the forces are
concentrated on the cylindrical belt. We gave the explanation of this
phenomenon based on torque balance on the particles and the nature of the
contacts. Finally we expressed the first and second normal stress differences
in terms of average particle level quantities, which explained some of the
properties of the normal stress differences.  One particular
non-monotonic behavior, i.e., how $N_1$ depends on the particle friction
$\mu_p$ for elongated particles, can be explained by the interplay between the
amount of velocity fluctuations and the orientation of the particles: 
large velocity fluctuations (collisional regime) or large Coulomb cone
with weak orientational order (collisional-frictional regime) increase the
isotropy of the force network, resulting in small values of $N_1$; while its
value in the intermediate (sliding friction) regime is high.

This work opens towards the rheology of elongated particles with more
complicated shapes or fibers with some flexibility, for which entanglement
effects are enhanced\cite{rodney-prl-2005,
bertails-descoubes-acmtransgraph-2011, gravish-chapter-2016}. Also, in the
context of active matter, it occurs that swimmers or bacteria can present an
elongated shape, which matters for their behaviour\cite{ilkanaiv-prl-2017}.
Beyond the properties of clustering of self-propelled
rods\cite{peruani-pre-2006, kudrolli-prl-2008, yang-pre-2010}, the extension
of the rheology of active dense granular flows\cite{peshkov-epl-2016} for such
long particles remains to be studied. Finally, the limitations of the $\mu(I)$
rheology have been recently emphasised, especially in the presence of strong
gradients with non-local effects coming into play\cite{
pouliquen-philtransrsoca-2009, kamrin-prl-2012, henann-pnas-2013,
bouzid-prl-2013, henann-prl-2014, kamrin-computpartmech-2014,
kamrin-softmatter-2015, bouzid-epl-2015,
bouzid-eurphysje-2015, rognon-jfluidmech-2015, kharel-epl-2018,
tang-softmatter-2018}, or as a source of ill-posedness in time dependent
calculations\cite{ barker-jfluidmech-2015, barker-jfluidmech-2017,
barker-procrsoca-2017, heyman-jfluidmech-2017, goddard-physfluids-2018}.
Because elongated particles can develop secondary flows and consequently build
gradients over time, these issues become crucial for the description of their
flows\cite{wortel-softmatter-2015}.

\section*{Acknowledgements}

This work was supported in part by the Hungarian National Research,
Development and Innovation Office NKFIH under grant OTKA K 116036.
We acknowledge funding from the CNRS with the PICS Grant No.\ 08187 `Flow properties
of granular materials consisting of elongated grains' (2019-2021),
and the Hungarian Academy of Sciences (Grant No.\ NKM-102/2019).
DBN was supported by the student scholarship of the French Embassy in
Budapest and Campus France.
We acknowledge KIF\"U for awarding us access to computational resources based
in Hungary at Debrecen.

\section*{References}

\bibliographystyle{iopart-num}
\bibliography{mui} 

\newpage

\pagestyle{empty}

\vspace*{-40mm}\hspace*{-37mm}\includegraphics[width=210mm]{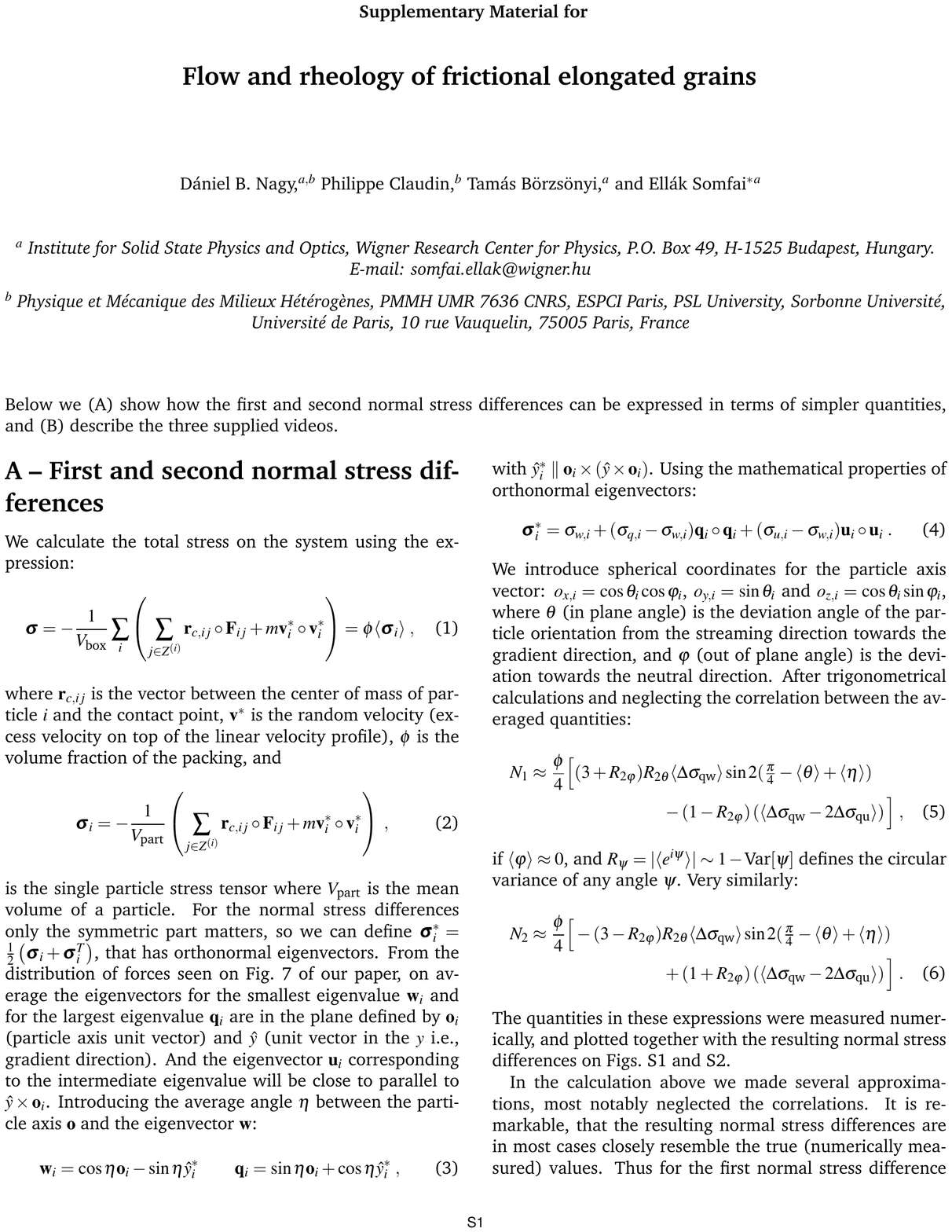}  %

\newpage
\vspace*{-40mm}\hspace*{-37mm}\includegraphics[width=210mm]{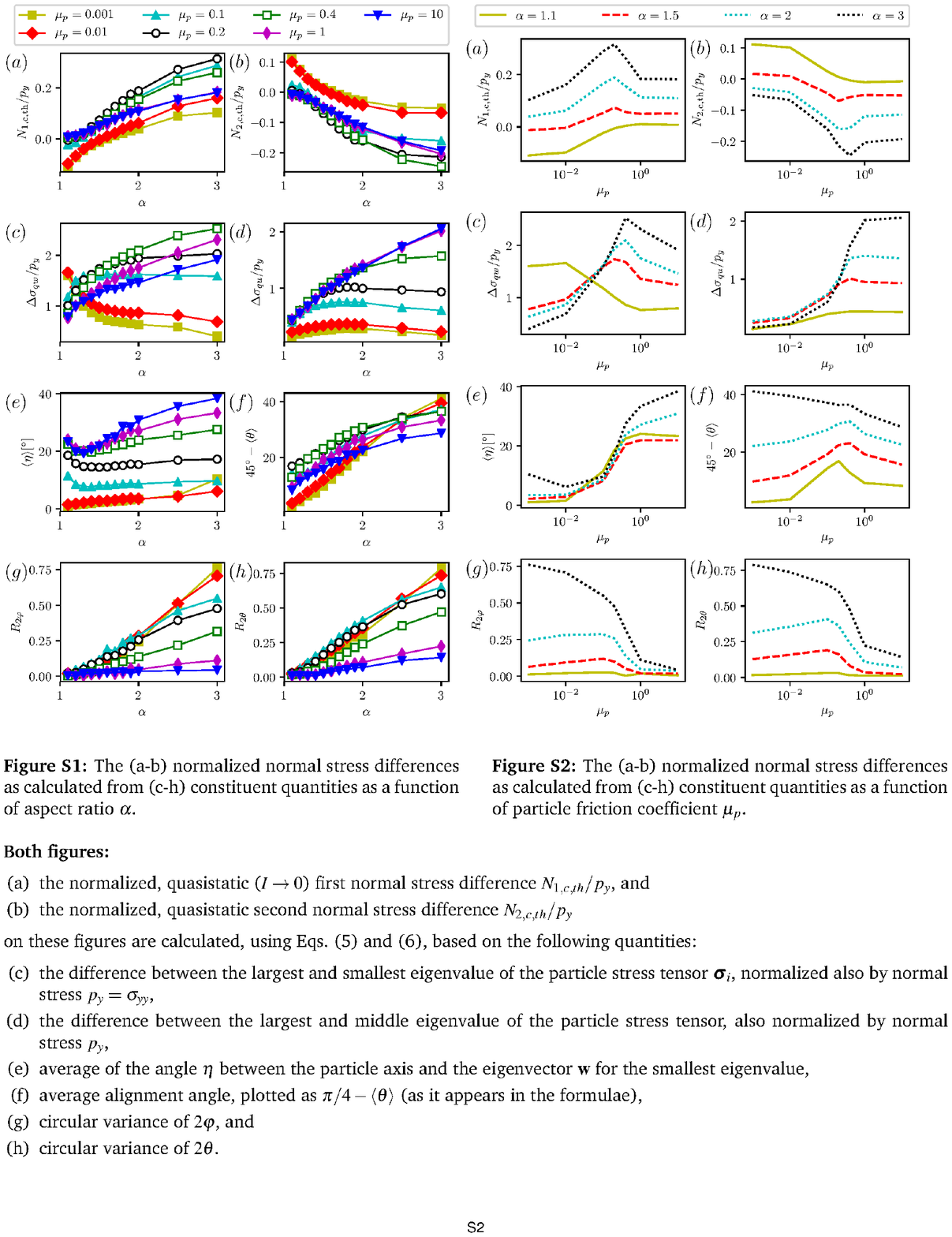}  %

\newpage
\vspace*{-40mm}\hspace*{-37mm}\includegraphics[width=210mm]{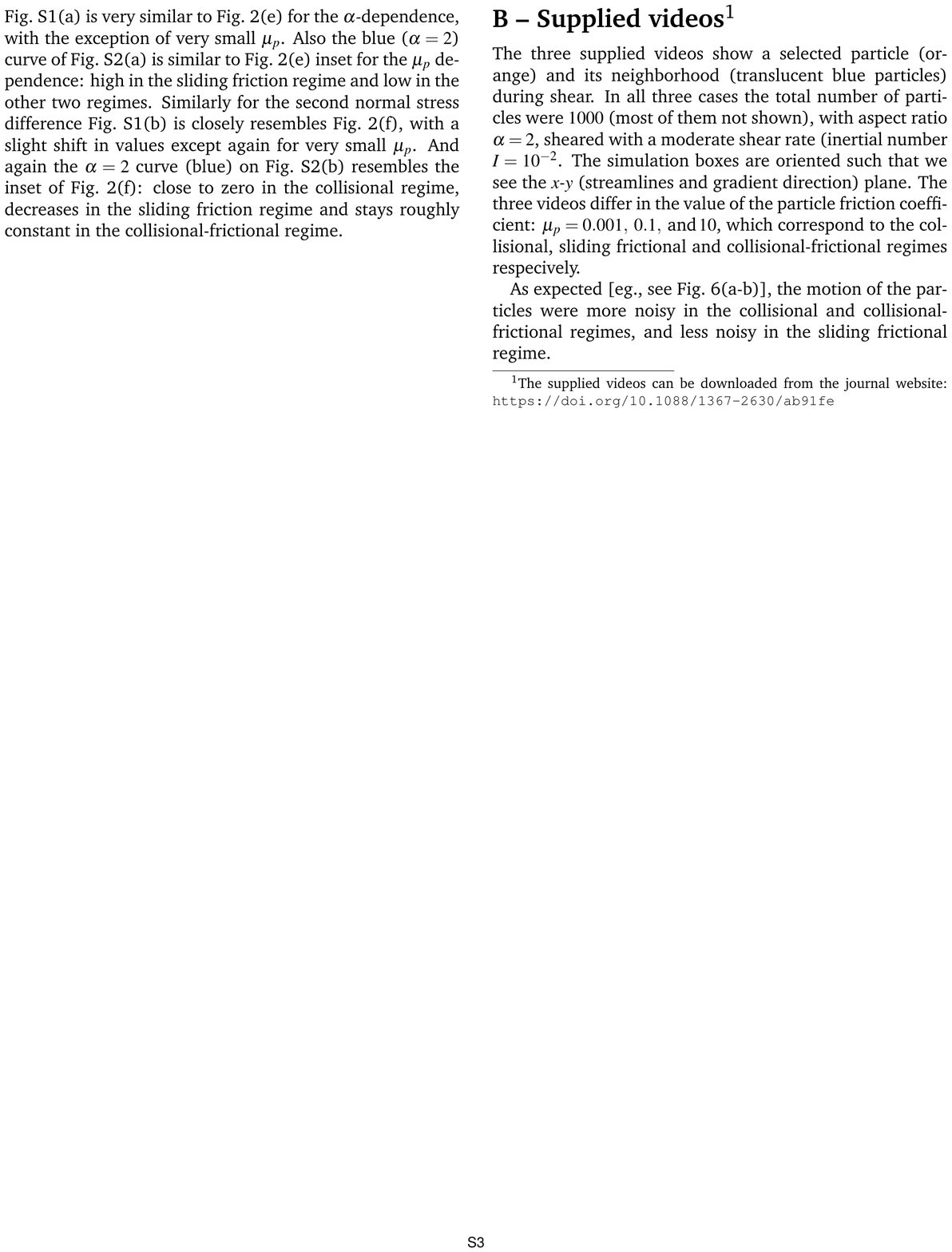}  %

\end{document}